\documentclass[aps,prb,showpacs,twocolumn,amsmath,amssymb,floatfix]{revtex4}
\usepackage{graphicx}
\usepackage{Jonasmacros}
\usepackage{bm}

\newcommand{\beqa}{\begin{eqnarray}}
\newcommand{\eeqa}{\end{eqnarray}}

\begin{document}
\preprint{}
\title{Detection of spin reversal and nutations through current measurements}
\author{Jonas Fransson}
\email{Jonas.Fransson@fysik.uu.se}
\affiliation{Department of Physics and Materials Science, Uppsala University, Box 530, 751 21\ \ UPPSALA, Sweden}
\begin{abstract}
The dynamics of a single spin embedded in a the tunnel junction between ferromagnetic contacts is strongly affected by the 
exchange coupling to the tunneling electrons. Moment reversal of the local spin induced by the bias voltage across the junction is shown to have a measurable effect on the tunneling current. Furthermore, the frequency of a harmonic bias voltage is picked up by the local spin dynamics and transferred back to the current generating a double frequency component.
\end{abstract}
\date{\today}
\maketitle

Detection and manipulation of single spins is an important field of science since it pushes the limits of quantum measurements. 
Single spins would also be objects for qubits and, thus, be crucial for quantum information technology. Potentially, spintronics will 
replace conventional electronics devices with spin analogues where manipulation, control, and read-out of spins enable 
functionality with no or little charge dynamics.\cite{awschalom2002} One main issue concerning measurements on the quantum 
limit is whether the local dynamics of quantum object generate measurable responses to externally applied fields.

Current-induced magnetic switching has been addressed for macroscopic planar\cite{slonszewski1996,berger1996,krivorotiov2005,edwards2005,nozaki2006} and magneto-mechanical\cite{kovalev2005} systems, and spin-dependent tunneling in systems with non-collinear configured ferromagnetic leads,\cite{braun2004,franssonPRB2005,weymann2007} as well as questions concerning spin-transfer torque,\cite{liu2003,bauer2003,ji2003,ozyilmaz2003} decoherence,\cite{katsura2006} and spin nutations
\cite{zhu2004,nussinov2005,zhu2006,fransson12007} in microscopic spin systems. It was recently demonstrated that the nutations of a local spin 
embedded in the tunnel junction between ferromagnetic leads can be electrically controlled and undergo stimulated spin reversal, 
using short bias voltage pulses.\cite{fransson22007}

\begin{figure}[b]
\begin{center}
\includegraphics[width=6.5cm]{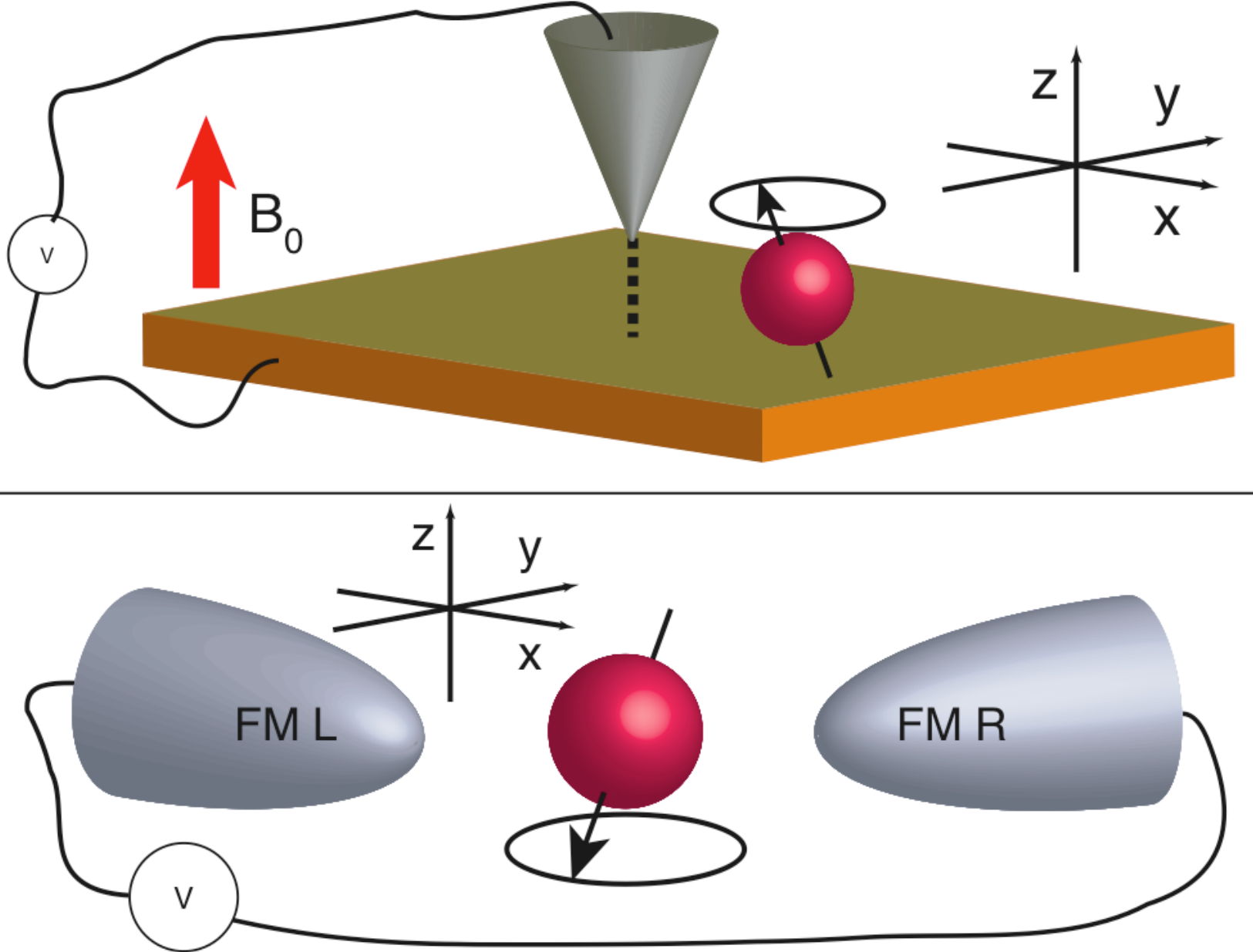}
\end{center}
\caption{(Color online) Schematic views of two possible set-ups relevant to the present study. In the upper panel, the local spin is located on a substrate surface for SP-STM measurement. The lower panel shows the local spin embedded in the tunnel junction between two ferromagnetic leads.}
\label{fig-system}
\end{figure}

The purpose of this paper is to discuss the effect of the local spin dynamics on the current flow through the system. In an 
arrangement with one ferromagnetic and one non-magnetic lead, the local spin tends to align with the magnetic moment of the 
ferromagnetic leads when this lead acts as the source, see Fig. \ref{fig-system} upper panel. This alignment leads to an enhanced tunneling through the junction since the local spin can mediate the tunneling electrons at a higher rate. If the ferromagnetic lead acts as the drain, however, the local spin align anti-parallel to the magnetic moment of the lead which decreases the tunneling through the junction. The local spin motion generates significant variations in the current. Second, we discuss the effect of a harmonic bias on systems arranged with two ferromagnetic leads in a non-parallel configuration, see Fig. \ref{fig-system} lower panel. The frequency $\omega_0$ of the bias voltage is picked up by the local spin motion and is transferred back to the current, generating current components with frequencies $\omega_0$ and $2\omega_0$. The doubled frequency component is a fingerprint of the interaction between the tunneling electrons and the local spin in presence of the bias induced magnetic field.

The back action of the tunnel current on the local spin and the resulting spin dynamics was studied in Ref. \onlinecite{fransson12007}.  Here, we implement the spin dynamics of the localized spin into the current and explore the prospect of detecting the spin  dynamics in the current. Consider a single spin embedded in a tunnel junction between two ferromagnetic leads, see Fig. \ref{fig-system}. The Hamiltonian  for the system can be written as
\begin{equation}
\Hamil=\Hamil_L+\Hamil_R+\Hamil_S+\Hamil_T,
\end{equation}
where $\Hamil_{L(R)}=\sum_{p(q)\sigma}\leade{p(q)}\cdagger{p(q)}\cc{p(q)}$, where $\cdagger{p(q)}\ (\cc{p(q)})$ creates 
(annihilates) an electron in the left (right) lead at the energy $\leade{p(q)}$ and spin $\sigma=\up,\down$. The spin Hamiltonian is 
simply $\Hamil_S=-g\mu_B\bfB_0\cdot\bfS$, where $g$ and $\mu_B$ is the gyromagnetic ratio and Bohr magneton, respectively, 
whereas $\bfB_0$ is the external magnetic field acting on the local spin $\bfS$. Finally, the tunneling Hamiltonian $\Hamil_T=
\sum_{pq\alpha\beta}(\csdagger{p\alpha}\hat{T}_{\alpha\beta}\cs{q\beta}+H.c.)$, where $\alpha,\beta$ are spin indices whereas 
the tunneling operator $\hat{T}_{\alpha\beta}=T_0\delta_{\alpha\beta}+T_1\bfS\cdot\bfsigma_{\alpha\beta}$, with the tunneling 
rate $T_0$ between the leads, the tunneling rate $T_1$ between the leads and the spin, and the vector of Pauli spin matrices $\bfsigma_{\alpha\beta}$.

Whenever there is a voltage applied across the junction, the tunneling current is given by $I(t)=-e\ddtinline\av{N_L}=ie
\av{\com{N_L}{\Hamil_T}}$, which results in $(\hbar=1$)
\begin{eqnarray}
I(t)&=&2e\re\sum_{\alpha\beta}
	\int_{-\infty}^te^{i\delta\phi(t,t')}
\nonumber\\&&\times
		\langle[\hat{T}_{\alpha\beta}(t)A_{\alpha\beta}(t),
		\hat{T}_{\beta\alpha}(t')A^\dagger_{\alpha\beta}(t')]\rangle
		dt',
\label{eq-current}
\end{eqnarray}
where the operator $A_{\alpha\beta}(t)=\sum_{pq}\csdagger{p\alpha}(t)\cs{q\beta}(t)$. Here, the transformed electronic 
annihilation operators
\begin{equation}
\cs{p(q)\alpha}(t)=e^{iK_{L(R)}t}\cs{p(q)\alpha}e^{-iK_{L(R)}t},
\end{equation}
with $K_{L(R)}=\Hamil_{L(R)}+(\mu_{L(R)}+W_{L(R)}(t))N_{L(R)}$ and $N_{L(R)}=\sum_{p(q)\sigma}\cdagger{p(q)}\cc{p(q)}$. The 
unequal chemical potentials of the system leads to the bias voltage $\mu_L+W_L(t)-\mu_R-W_R(t)=e[V_\text{dc}+V_\text{ac}(t)]$, 
and we have defined $\phi(t)=e\int_{-\infty}^t[V_\text{dc}+V_\text{ac}(t')]dt'$, and $\delta\phi(t,t')=\phi(t)-\phi(t')$. Here we will 
consider biasing of the system with 
\begin{equation}
\delta\phi_1(t,t')=
	e\int_{t'}^t(V_{dc}+V_{ac}[\Theta(\tau-\tau_0)-\Theta(\tau-\tau_1)])d\tau,
\label{eq-V1}
\end{equation}
with $\tau_0<\tau_1$, and
\begin{equation}
\delta\phi_2(t,t')=
	e\int_{t'}^t(V_{dc}+V_{ac}\cos{\omega_0\tau})d\tau,
\label{eq-V2}
\end{equation}
For later reference, we also define $\xi_{p(q)\sigma}=\leade{p(q)}-\mu_{L(R)}$. 

Before deriving the expression for the current we briefly discuss the dynamics of the local spin, for which an equation was derived in Ref. 
\onlinecite{fransson12007} and found to be well described by the Landau-Lifshitz-Gilbert equation\cite{landau1935,gilbert1955}
\begin{equation}
\frac{d\bfS(t)}{dt}=\alpha(t)\frac{d\bfS(t)}{dt}\times\bfS(t)+g\mu_B\bfS(t)\times\bfB(t),
\label{eq-llg}
\end{equation}
where $\bfB=\bfB_0+\bfB_\text{ind}^{(1)}+\bfB_\text{ind}^{(2)}$ is the effective magnetic field. The energies 
associated with the spin dynamics, $\hbar\omega_L\sim1\ \mu$eV, are much smaller than the energies for the electronic 
processes which are of the order of at least 1 meV. The damping factor $\alpha(t)\sim T_1^2/D$ at low temperatures, where $2D$ is the band widths in of the leads,\cite{fransson12007} and it is reasonable to believe that this is the case also for finite temperatures. The leads are assumed to have large $D$, hence, $\alpha$ becomes negligible and as a result Eq. (\ref{eq-llg}) can be  analytically solved. 

The induced magnetic fields can be expressed as\cite{fransson12007,fransson22007}
\begin{eqnarray}
\lefteqn{
g\mu_B\bfB_\text{ind}^{(1)}(t)=
	-2T_0T_1\sum_{pq\sigma}\sigma_{\sigma\sigma}^z
		[f(\xi_{p\sigma})-f(\xi_{q\sigma})]
}
\nonumber\\&&\times
	\im\int_{-\infty}^te^{i[(\xi_{p\sigma}-\xi_{q\sigma})(t-t')+\delta\phi(t,t')]}dt'\hat{\bf z}
\label{eq-B1}
\end{eqnarray}
\begin{eqnarray}
g\mu_B\bfB_\text{ind}^{(2)}(t)&=&
	S\int_{-\infty}^t \Bigl\{
	K_{xy}^{(2)}(t,t')(n_y\hat{\bf x}-n_x\hat{\bf y})
\nonumber\\&&
	+[K_{xx}^{(2)}(t,t')-K_{zz}^{(2)}(t,t')]n_z\hat{\bf z}\Bigr\}dt'
\label{eq-B2}
\end{eqnarray}
where
\begin{eqnarray}
K_{xy}^{(2)}(t,t')&=&2T_1^2
	\re\sum_{pq\sigma}[f(\xi_{p\sigma})-f(\xi_{q\bar\sigma})]
\nonumber\\&&\times
			e^{i[(\xi_{p\sigma}-\xi_{q\bar\sigma})(t-t')+\delta\phi(t,t')]}
\label{eq-Kxy}
\end{eqnarray}
and 
\begin{eqnarray}
\lefteqn{
K_{xx}^{(2)}(t,t')-K_{zz}^{(2)}(t,t')=}
\nonumber\\&&
	-2T_1^2\im\sum_{pq\sigma}
			\Bigl\{[f(\xi_{p\sigma})-f(\xi_{q\bar\sigma})]
			e^{i(\xi_{p\sigma}-\xi_{q\bar\sigma})(t-t')}
\nonumber\\&&
	-[f(\xi_{p\bar\sigma})-f(\xi_{q\sigma})]
			e^{i(\xi_{p\bar\sigma}-\xi_{q\sigma})(t-t')}\Bigr\}e^{i\delta\phi(t,t')}
\label{eq-Kxz}
\end{eqnarray}

The local spin-dynamics can be described by its corresponding classical motion, which is consistent with the parametrization $
\bfS=S{\bf n}=S(\cos\varphi\sin\theta,\sin\varphi\sin\theta,\cos\theta)$. Assuming time-independent external magnetic field $
\bfB_0=B_0\hat{\bf z}$, results in the equations of motion
\begin{equation}
\begin{array}{rcl}
\displaystyle \frac{d\varphi}{dt} & = &
	\displaystyle - g\mu_B[B_0+B_\text{ind}^{(1)}(t)]
\\& &	\displaystyle
	+S\int_{-\infty}^t[K_{xx}^{(2)}(t,t')-K_{zz}^{(2)}(t,t')]dt'\cos{\theta},
\\\\
\displaystyle \frac{d\theta}{dt} & = &
	\displaystyle S\int_{-\infty}^t K_{xy}^{(2)}(t,t')dt'\sin{\theta}.
\end{array}
\label{eq-angles}
\end{equation}

The expression for the current is conveniently divided into three components, e.g. $I(t)=I_0(t)+I_z(t)+I_\perp(t)$, and using that the 
local spin is semi-classically described these contributions can be written as
\begin{eqnarray}
I_0(t)&=&2e\re\sum_{pq\sigma}\int_{-\infty}^t
	T_0^2[f(\xi_{p\sigma})-f(\xi_{q\sigma})]
\nonumber\\&&\times\vphantom{\biggl(}
	e^{i(\xi_{p\sigma}-\xi_{q\sigma})(t-t')+i\delta\phi(t,t')}dt'
\end{eqnarray}
which accounts for the direct tunneling between the leads, while
\begin{eqnarray}
I_z(t)&=&4e\re\sum_{pq\sigma}\int_{-\infty}^t
	T_0T_1\biggl[\sigma_{\sigma\sigma}^z
		+\frac{T_1}{2T_0}\cos{\theta(t)}\biggr]\cos{\theta(t)}
\nonumber\\&&\hspace{-0.5cm}\times\vphantom{\biggl(}
	[f(\xi_{p\sigma})-f(\xi_{q\sigma})]
	e^{i(\xi_{p\sigma}-\xi_{q\sigma})(t-t')+i\delta\phi(t,t')}dt',
\end{eqnarray}
is influenced by the local spin moment, and
\begin{eqnarray}
I_\perp(t)&=&2e\re\sum_{pq\sigma}\int_{-\infty}^t
	T_1^2\sin^2{\theta(t)}
		[f(\xi_{p\sigma})-f(\xi_{q\bar\sigma})]
\nonumber\\&&\times\vphantom{\biggl(}
		e^{i(\xi_{p\sigma}-\xi_{q\bar\sigma})(t-t')+i\delta\phi(t,t')}dt',
\end{eqnarray}
includes spin-flip of the tunneling electron caused by the presence of the local spin. The resulting current is clearly modulated by  the presence of the spin 
through the variation of the spin polar angle $\theta(t)$, and through its doubled frequency $2\theta(t)$.  This property will be exploited later in this paper, and it 
will be shown that the nutations of the local spin which are induced by the bias voltage can be measured in the current as higher order harmonics. The expressions for the current also show that the azimuthal motion, $\varphi(t)$, has no direct influence on the current. This would also be expected for a spin influenced only by an external magnetic field $\bfB_0=B_0\hat{\bf z}$. Thus, we need only to determine the polar angle motion of the local spin, for a 
complete description of its effect on the current. The general solution for the polar angle motion can be written
\begin{equation}
\theta(t)=2\arctan\biggl(\tan{\frac{\theta_0}{2}}
	e^{S\int_{t_0}^t\int_{-\infty}^{t'}K_{xy}^{(2)}(t',t'')dt''dt'}\biggr),
\label{eq-gentheta}
\end{equation}
where $\theta_0$ is the initial polar angle at the initial time $t_0$.

For the bias $\delta\phi_1(t,t')$, Eq. (\ref{eq-V1}), we have
\begin{eqnarray}
\frac{d\theta(t)}{dt}&=&-2\pi ST_1^2
	\sum_\sigma\sigma_{\sigma\sigma}^zN_{L\sigma}N_{R\bar\sigma}
	\biggl\{eV_{dc}\Theta(\tau_0-t)
\nonumber\\&&
	+\biggl[eV_{dc}\cos{[eV_{ac}(t-\tau_0)]}e^{-(t-\tau_0)/\tau}
\nonumber\\&&
	+e(V_{dc}+V_{ac})
	\biggl(1-e^{-(t-\tau_0)/\tau}\biggr)\biggr]
	[\Theta(t-\tau_0)
\nonumber\\&&
	-\Theta(t-\tau_1)]
	+\biggl[eV_{dc}\cos{[eV_{ac}(\tau_1-\tau_0)]}
\nonumber\\&&\times
	e^{-(t-\tau_0)/\tau}
	+e(V_{dc}+V_{ac})\biggl(e^{-(t-\tau_1)/\tau}
\nonumber\\&&
	-e^{-(t-\tau_0)/\tau}\biggr)
	\cos{[eV_{ac}(t-\tau_1)]}
	+eV_{dc}\biggl(1
\nonumber\\&&
	-e^{-(t-\tau_1)/\tau}\biggr)\biggr]
	\Theta(t-\tau_1)\biggr\}\sin{\theta(t)},
\label{eq-theta}
\end{eqnarray}
where the time-scale $\tau$ relates to the electronic tunneling processes. Such processes are of the order 1 fs, which is much 
faster than the characteristic time-scale for the spin dynamics for biases ranging in 1 | 100 mV. For e.g. a 1 ns bias pulse of 1 mV, the critical time scale is $\sim100$ fs, which should be within the realms of present experimental state-of-the-art-technology for nanoscale systems. Physically, the time-scale $\tau$ means that the induced magnetic field is a retarded response to the bias across the junction. In Eq. (\ref{eq-theta}), $N_{L(R)\sigma}$ is the spin $\sigma$ density of  states (DOS) in the left (right) lead.

The characteristic time-scale $\tau_c$ for the polar angle motion can be written as $1/\tau_c\simeq2\pi eVT_1^2\sum_\sigma
\sigma_{\sigma\sigma}^zN_{L\sigma}N_{R\bar\sigma}$. Parametrizing the spin-polarized DOS $N_{L(R)\sigma}=N_0(1+\sigma 
p_{L(R)})/2$,\cite{fransson2005} where $-1\leq p_{L(R)}\leq1$, assuming $T_1N_0\sim0.1$, $p_L=-p_R=1/2$, and $V\sim1$ mV, 
gives $\tau_c\simeq5$ ps, which is sufficiently short to switch the spin from being $\down$ to $\up$ within 1 ns.

\begin{figure}[t]
\begin{center}
\includegraphics[width=7.5cm]{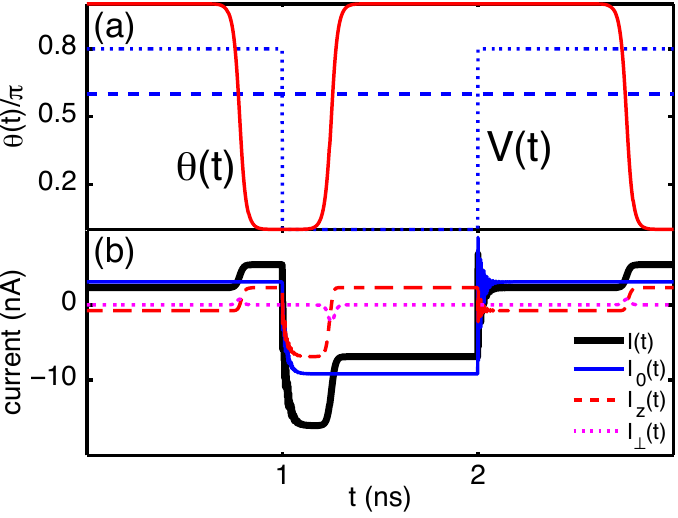}
\end{center}
\caption{(Color online) a) Time-dependence of the polar angle $\theta(t)$ (solid) for the pulsed bias voltage $V(t)=eV_{dc}
+eV_{ac}[\Theta(t-\tau_0)-\Theta(t-\tau_1)]$, $(\tau_0,\tau_1)=(1,2)$ ns (dotted) (zero bias indicated through dashed line). b) Time-
dependence of the total current $I(t)$ (bold), and the contributions $I_0(t)$ (faint), $I_z(t)$ (dashed), and $I_\perp(t)$ (dotted).}
\label{fig-Ipulse}
\end{figure}
Assuming initially that $\theta_0=\pi$, that is, the local spin moment is $\down$, see Fig. \ref{fig-Ipulse} a), showing the time-dependence of the polar angle for the bias pulse corresponding to $\delta\phi_1(t,t')$. Also, assume that $V_{dc}>0$, and that 
$p_L=1/2$, $p_R=0$, for $T_1N_0\sim0.1$. The assumption is thus that one lead is ferromagnetic whereas the other is non-magnetic, which would be applicable to scanning tunneling microscopy (STM) measurements with a spin-polarized tip. The 
contribution from $I_z$ is then finite whereas $I_\perp=0$, see Fig. \ref{fig-Ipulse} b) showing the total current $I(t)$ (bold), 
and the contributions $I_0(t)$ (faint), $I_z(t)$ (dashed), and $I_\perp(t)$ (dotted). The influence of the stationary field tends to flip 
the spin moment to $\up$,\cite{fransson22007} which can be understood from Eq. (\ref{eq-gentheta}) where the exponent has a 
negative sign. The contribution from $I_z$ increases at the instant of the spin-flip, since the local spin then aligns with the majority 
spin in the source (left lead) and can thus mediate the tunneling electrons at a higher rate. Momentarily, as the local spin flips from 
$\down$ to $\up$, the polar angle $\theta(t)\neq0,\pm\pi$ which implies that the contribution from $I_\perp$ peaks during a short 
time period. The peak of this contribution will, however, be hidden by the sharp increase of $I_z$ during the same time period. 
Nonetheless, this analysis shows that the total current through the system increases as a response to the spin-flip of the local spin, 
see Fig. \ref{fig-Ipulse} b) (bold).

\begin{figure}[t]
\begin{center}
\includegraphics[width=7.5cm]{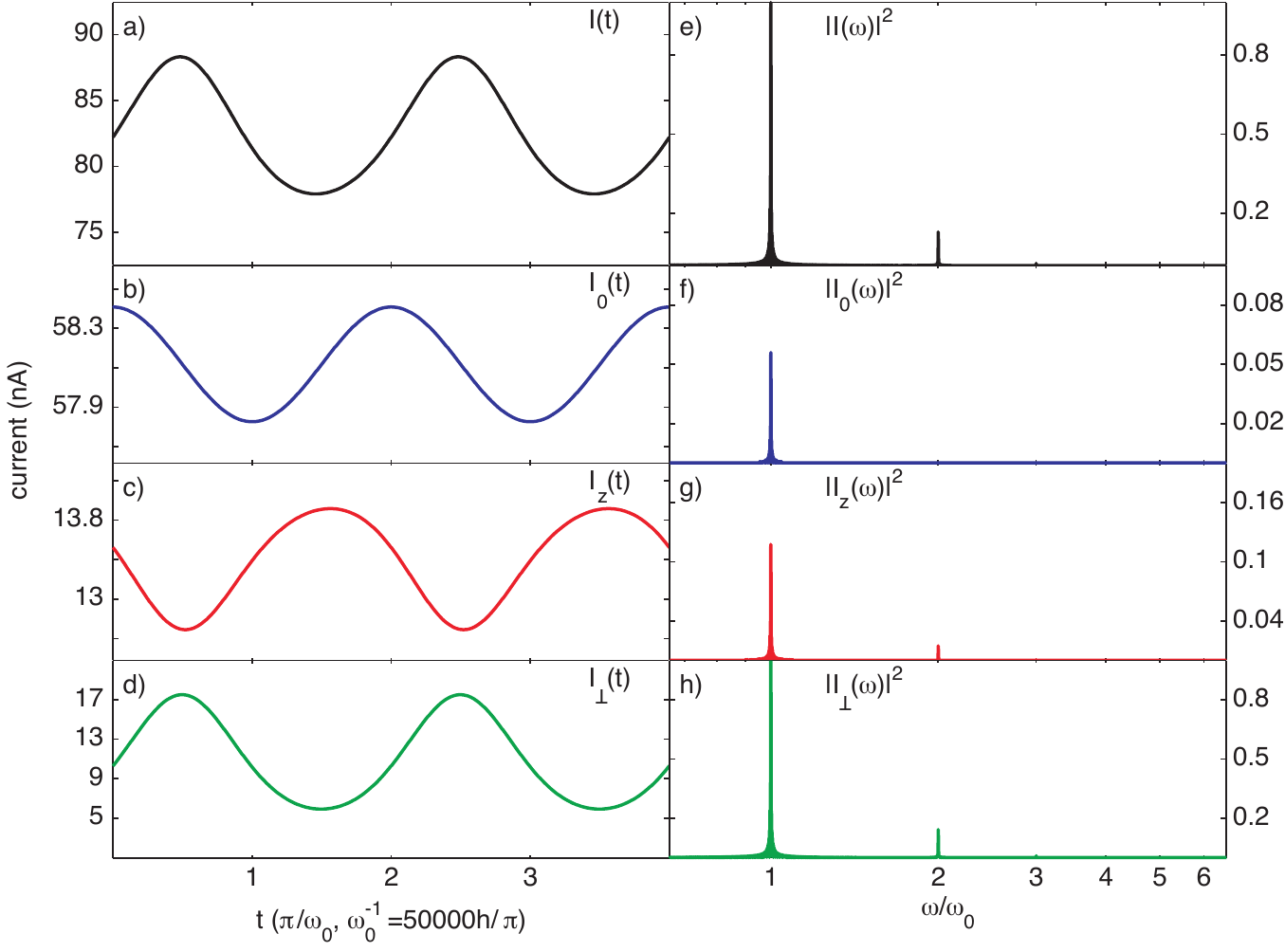}
\end{center}
\caption{(Color online) a) Total current $I(t)$ through the system biased with $V(t)=V_{dc}+V_{ac}\cos\omega_0t$, $
\omega_0^{-1}=5h\cdot10^4/\pi$, and e) its Fourier transform showing a main peak at $\omega=\omega_0$ and a 
second peak at $\omega=2\omega_0$. b) The current contribution $I_0(t)$, and f) its Fourier transform showing a single peak at $
\omega=\omega_0$. c), d) The current contributions $I_z(t)$ and $I_\perp(t)$, respectively, and g), h) their corresponding Fourier 
transforms, both showing peaks at $\omega=\omega_0,\ 2\omega_0$. Panels e) - h) display the relative intensity of the 
contributions in arbitrary units. Here, the spin-polarization $p_L=-p_R=0.9$, $T_1/T_0=1/2$, $N_0T_1\sim0.1$, $eV_{dc}
\sim1000\omega_0$, and $eV_{ac}\sim5\omega_0$.}
\label{fig-I}
\end{figure}

Now, assume that the amplitude of the bias pulse $V_{ac}<-V_{dc}<0$. This bias pulse causes the current to flow from the right to 
left, and the induced magnetic field, which affects the polar angle motion of the spin, works as to reverse the moment of local spin. 
This can be seen in the sign of the exponent in Eq. (\ref{eq-gentheta}), which is positive and, thus, signifies that the polar angle approaches $
\pi$. Thus, we assume that the bias pulse is sufficiently strong and/or long to stimulate the spin reversal (that is going from $\up$ to $
\down$), see Fig. \ref{fig-Ipulse} a). At the instant of the spin-flip, the contribution from $I_z$ decreases since the local spin moment 
becomes opposite to spin-polarization of the drain (left lead), see Fig. \ref{fig-Ipulse} b) (dashed). The local spin mediates the 
tunneling electron at a slower rate. Analogously as in the above discussion, the contribution from $I_\perp$ peaks during the short 
time period of the spin reversal, see Fig. \ref{fig-Ipulse} b) (dotted), nevertheless, the net effect of the spin-flip is a reduced 
amplitude of the current, see Fig. \ref{fig-Ipulse} b) (bold). As the pulse terminates, the bias again becomes positive which reverses 
the local spin moment from $\down$ to $\up$. It should also be noticed that while the contribution from $I_0(t)$ follows the 
biasing of the system, it is unaffected by the local spin dynamics, as expected, see Fig. \ref{fig-Ipulse} b) (faint).

For a harmonic bias $\delta\phi_2(t,t')$, the current response to the local spin displays higher order harmonics which are only 
present when the leads are ferromagnetic and in a non-parallel configuration. The harmonic modulation, with frequency $\theta(t)$, of 
the bias is picked up by the local spin, however, its presence is transferred to the current such that a second harmonic, $2\theta(t)
$, appears in the current modulation. This can be directly seen in the expressions for the currents $I_z$ and $I_\perp$, both of which 
have terms with the factors $\cos2\theta(t)$.

The transverse induced magnetic field $\int_{-\infty}^tK_{xy}^{(2)}dt'$, and the currents $I_z$ and $I_\perp$ all have the time-
dependence $(x=\xi_{p\sigma}-\xi_{q\sigma'})$
\begin{eqnarray}
\lefteqn{
	\int_{-\infty}^t
	e^{i(x+eV_{dc})(t-t')+ieV_{ac}(\sin\omega_0t-\sin\omega_0t')/\omega_0}dt'\approx
}
\nonumber\\&&
	\pi\sum_nJ_n(eV_{ac}/\omega_0)
		J_m(eV_{ac}/\omega_0)\delta(x+eV_{dc}+n\omega_0)
\nonumber\\&&\times
		e^{i\omega_0(n-m)t}.
\label{eq-tharm}
\end{eqnarray}
It should be noticed that although this time-dependence arise in the current contribution $I_0(t)$, the net effect of the bias still 
yields a single harmonic variation of this contribution, see Fig. \ref{fig-I} b) and f), illustrating the time-dependence of $I_0(t)$ 
and its Fourier transform. The plots clearly shows that the direct tunneling current  acquires only the frequency of the bias and no 
higher harmonics. This can be understood from the fact that $I_0(t)$ is a direct kinetic response to the electric field and does not 
reflect any of the local (spin) dynamics in the tunnel junction.

\begin{figure}[t]
\begin{center}
\includegraphics[width=7.5cm]{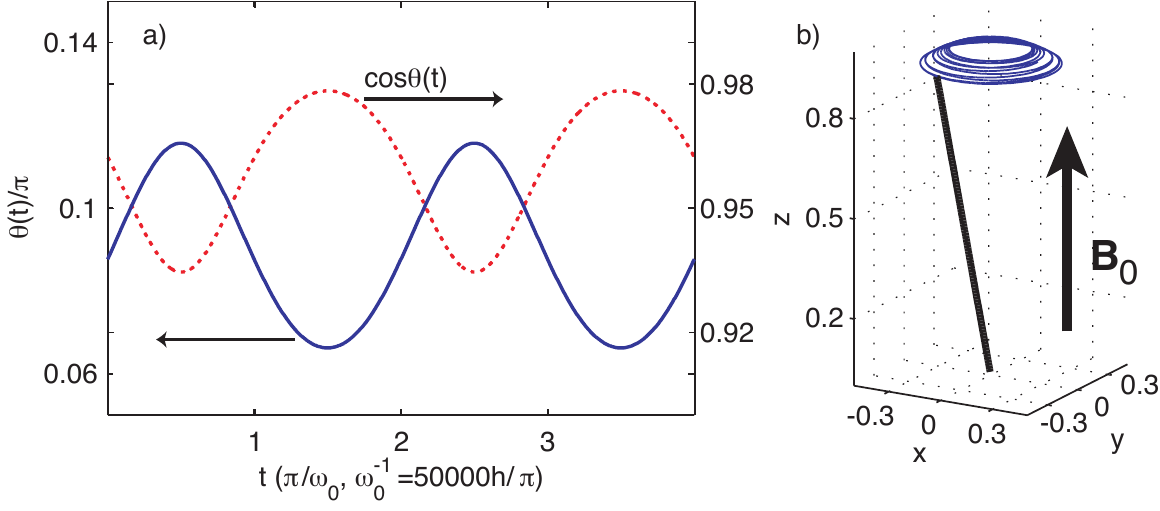}
\end{center}
\caption{(Color online) a) Polar angle motion (solid) and $\cos\theta(t)$ (dotted) for the bias $V(t)=V_{dc}+V_{ac}\cos\omega_0t$, and b) spin motion. In panel b), an external magnetic field $\bfB_0=1$ T has been added in order to speed up the azimuthal rotation, whereas other parameters are as in Fig. \ref{fig-I}.}
\label{fig-theta}
\end{figure}

The polar angle motion resulting from the time-dependence in Eq. (\ref{eq-tharm}) is given by
\begin{eqnarray}
\theta(t)&=&2\arctan\biggl(\tan\frac{\theta_0}{2}
	\prod_{nm}\exp\biggl[2\pi ST_1^2
\nonumber\\&&\times
		\sum_\sigma N_{L\sigma}N_{R\bar\sigma}\sigma_{\sigma\sigma}^z
		J_n(eV_{ac}/\omega_0)J_m(eV_{ac}/\omega_0)
\nonumber\\&&\times
		(eV_{dc}+n\omega_0)\frac{\sin\omega_0(n-m)t}{\omega_0(n-m)}\biggr]\biggr),
\end{eqnarray}
showing that the nutations of the spin caused by the bias induced magnetic field are directly modulated by the harmonic bias. 
Moreover, from this expression of the polar angle motion, it is clear that the lower the frequency, i.e. $\omega_0\ll eV_{dc}/10$,  the 
larger the amplitude of the angular displacement. Therefore, low frequencies provide a greater response to the current 
since it is influenced by the spin nutations through $\sin\theta(t)$ and $\cos\theta(t)$.

In order to acquire a large response from the spin nutations it is also required that the magnetic moments of the leads are non-parallel. Therefore, assume that $p_L=-p_R=0.9$. Moreover, in order to further amplify the response from the spin nutations, we 
assume that $T_1/T_0=1/2$, $N_0T_1\sim0.1$, and let $eV_{dc}\sim1000\omega_0$ and $eV_{ac}\sim5\omega_0$. Under those 
conditions, which are reasonable from an experimental view point, the amplitude of the polar angle $(\max\theta-\min\theta)/\pi
\sim0.05$, Fig. \ref{fig-theta} a) (solid), which is a sufficiently large displacement to generate a significant contribution from $\cos
\theta(t),\ \sin\theta(t)$, see Fig. \ref{fig-theta} a) (dotted), which is transferred into the current contributions $I_z(t)$ and $I_\perp(t)$.

\begin{figure}[t]
\begin{center}
\includegraphics[width=7.5cm]{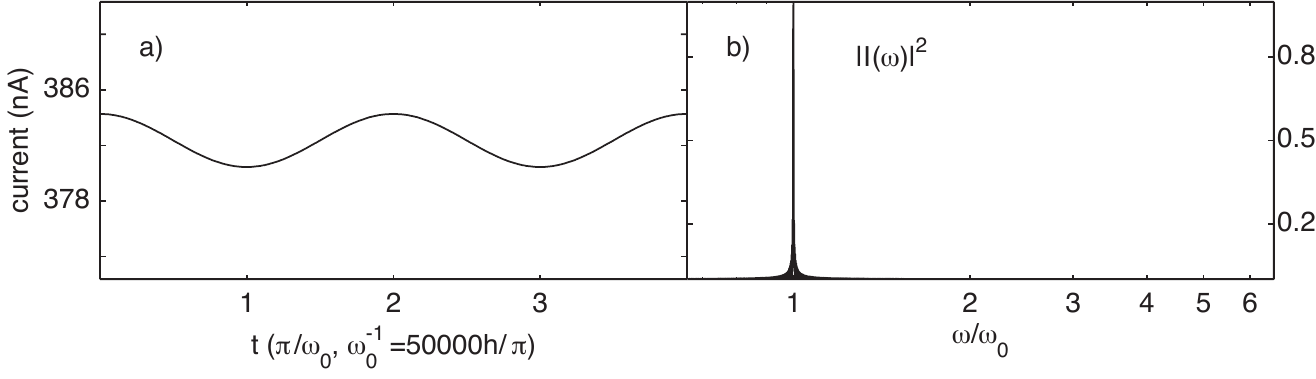}
\end{center}
\caption{Total a) time-dependent current $I(t)$, and b) its Fourier transform $|I(\omega)|$ in the case of non-spin-polarized leads, 
e.g. $p_L=p_R=0$. Other parameters are as in Fig. \ref{fig-I}.}
\label{fig-IB0}
\end{figure}
The current contributions $I_z(t)$ and $I_\perp(t)$, and their corresponding Fourier transforms, are plotted in Fig. \ref{fig-I} c), d), 
g), and h), which illustrate that their contributions modify the shape of the total current c.f. panels a), c), and d) in Fig. \ref{fig-I}, and 
that this modification arise from current contributions with frequency $2\omega_0$, c.f. panels e), g), and h). The contributions with 
this double frequency is a direct fingerprint of the (exchange) interaction between tunneling electrons and the local spin, in 
combination with the spin-polarization of the leads. In case of non-spin-polarized leads there is no magnetic field induced from the 
biasing of the system\cite{fransson12007} and, thus, the time-dependent modulation of the bias is not transferred to the local 
spin dynamics. Under those circumstances the current consist of a single frequency contribution, see Fig. \ref{fig-IB0} which 
illustrates the time-dependence a) and Fourier transform b) of the total current, which is understood since then the current is simply 
a kinetic response to the electric field with no influence from the local (spin) dynamics.

In summary, the effect of a local spin, embedded in the tunnel junction between ferromagnetic leads, on the current through the 
system has been studied. It has been shown that the spin reversal of the local spin, stimulated by the bias voltage induced 
magnetic field in the junction, generates a measurable variation of the current. This feature thus enable direct recording of the 
local spin reversal in the tunneling current. It has, moreover, been demonstrated that the spin nutations induced by a harmonic 
bias voltage with frequency $\omega_0$ is transferred back into the current, generating components with frequencies $
\omega_0$ and $2\omega_0$. The double frequency component is a direct response of the spin motion in the current, which 
vanishes whenever the transversal induced magnetic field is absent. It was also shown that the $2\omega_0$ component is 
sufficiently large to be measured within the realms of todays state-of-the-art nanotechnology. Experiments of the discussed spin 
dynamics measurements would be extremely intriguing and useful for further advances within basic science of nanoscale physics.

The author thanks J. -X. Zhu and A. V. Balatsky for useful discussions. Special acknowledgments to the genuine and professional scientific atmosphere provided at the International Ten Bar Caf\'e.

\end{document}